\documentclass[conference]{IEEEtran}
\IEEEoverridecommandlockouts
\usepackage{amsmath,amssymb,amsfonts}
\usepackage{algorithmic}
\usepackage{graphicx}
\usepackage{textcomp}
\usepackage{xcolor}
\usepackage{booktabs}
\def\BibTeX{{\rm B\kern-.05em{\sc i\kern-.025em b}\kern-.08em
    T\kern-.1667em\lower.7ex\hbox{E}\kern-.125emX}}

\usepackage[capitalize]{cleveref}
\usepackage{todonotes}

\usepackage[activate]{microtype}
\usepackage[natbib, bibencoding=utf8, citestyle=numeric, bibstyle=ieee, maxbibnames=999, maxcitenames=2, mincitenames=1, sortcites]{biblatex}
\bibliography{references}
\usepackage[acronym, shortcuts, nohypertypes={acronym}]{glossaries}
\newacronym{ADI}{ADI}{acoustic diversity index}
\newacronym{AI}{AI}{artificial intelligence}
\newacronym{CI}{CI}{confidence interval}
\newacronym{CNN}{CNN}{convolutional neural network}
\newacronym{DL}{DL}{deep learning}
\newacronym{DNN}{DNN}{deep neural network}
\newacronym{ML}{ML}{machine learning}
\newacronym{SDG}{SDG}{Sustainable Development Goal}
\newacronym{STFT}{STFT}{short-time Fourier transform}
\newacronym{NDSI}{NDSI}{normalised difference soundscape index}
\newacronym{UAR}{UAR}{unweighted average recall}
\newacronym{USV}{USV}{ultrasound vocalisation}

\begin{document}

\title{An automatic analysis of ultrasound vocalisations for the prediction of interaction context in captive Egyptian fruit bats
\thanks{This work was funded from the DFG project No.\ 512414116 (HearTheSpecies) and the DFG project No.\ 442218748 (AUDI0NOMOUS).}
}

\author{\IEEEauthorblockN{
Andreas Triantafyllopoulos\IEEEauthorrefmark{1}\IEEEauthorrefmark{2}\IEEEauthorrefmark{3},
Alexander Gebhard\IEEEauthorrefmark{1}\IEEEauthorrefmark{2}\IEEEauthorrefmark{3},
Manuel Milling\IEEEauthorrefmark{1}\IEEEauthorrefmark{2}\IEEEauthorrefmark{3},
Simon Rampp\IEEEauthorrefmark{2},
Bj\"orn Schuller\IEEEauthorrefmark{1}\IEEEauthorrefmark{2}\IEEEauthorrefmark{3}\IEEEauthorrefmark{4}\IEEEauthorrefmark{5},
}
\IEEEauthorblockA{\IEEEauthorrefmark{1}CHI -- Chair of Health Informatics, Technical University of Munich, MRI, Munich, Germany}
\IEEEauthorblockA{\IEEEauthorrefmark{2}EIHW -- Chair of Embedded Intelligence for Health Care and Wellbeing, Augsburg, Germany}
\IEEEauthorblockA{\IEEEauthorrefmark{3}MCML -- Munich Center for Machine Learning, Munich, Germany}
\IEEEauthorblockA{\IEEEauthorrefmark{4}MDSI -- Munich Data Science Institute, Munich, Germany}
\IEEEauthorblockA{\IEEEauthorrefmark{5}GLAM -- Group on Language, Audio, \& Music, Imperial College London, UK}
\IEEEauthorblockA{Email: andreas.triantafyllopoulos@tum.de}
}

\maketitle

\begin{abstract}

Prior work in computational bioacoustics has mostly focused on the detection of animal presence in a particular habitat. However, animal sounds contain much richer information than mere presence; among others, they encapsulate the interactions of those animals with other members of their species. Studying these interactions is almost impossible in a naturalistic setting, as the ground truth is often lacking. The use of animals in captivity instead offers a viable alternative pathway. However, most prior works follow a traditional, statistics-based approach to analysing interactions. In the present work, we go beyond this standard framework by attempting to predict the underlying context in interactions between captive \emph{Rousettus Aegyptiacus} using deep neural networks. We reach an unweighted average recall of over 30\% -- more than thrice the chance level -- and show error patterns that differ from our statistical analysis.
This work thus represents an important step towards the automatic analysis of states in animals from sound. 
\end{abstract}


\section{Introduction}

Among other signals and cues, several animals rely on the use of vocalisations to transmit information to other members of their species and beyond~\citep{Bradbury98-POA}.
The detection and subsequent analysis of those vocalisations falls under the premise of \emph{computational bioacoustics}~\citep{Stowell22-CBW, Schuller24-ECA}.
The field has its roots in \emph{ecoacoustics} -- the broader study of sounds and ecology~\citep{Farina17-ETE, Muller22-LUI}, which aims to establish a (causal) link between habitat structure and different ecological indicators on the composition of its soundscapes.
Traditionally, this has been done through the use of expert-based features~\citep{Farina17-ETE}.

By necessity, a large focus has been placed on \emph{detection} as the first step in any analysis pipeline~\citep{Rizos24-PVM}; this corresponds to identifying the presence of animals in soundscapes, often passively acquired by sensors placed in fixed positions and recording continuously.
Moreover, this is linked to a key promise of bioacoustics: the automatic, continuous monitoring of biodiversity.
This promise can be a key facilitator for two of the UN's 17 \acp{SDG}, \emph{Life Below Water} and \emph{Life on Land}, as monitoring biodiversity is a necessary prerequisite for preserving it and understanding the (positive or negative) impacts that any human interventions may have.
While the field is still struggling with challenges related to the lack of annotated data~\citep{Gebhard24-EMI} -- especially for rare species -- and generalisation across different landscapes, the use of \acp{DNN} has led to much progress on achieving this goal.

A parallel focus lies on the \emph{understanding} of the underlying meaning encompassed in those signals.
Oftentimes animal sounds encode a direct message, such as an alarm or flight call~\citep{Bradbury98-POA}.
Distinguishing between these different messages can help interpret animal behaviour.
However, vocalisations may also encode information about the emitter's health status, its social standing in a tribe hierarchy, its reproductive intent, and other attributes that characterise a particular animal at a specific point in time.
Here, the field is arguably in more nascent stages, especially when we compare it to the analysis of such states and traits in humans, where tools and available technologies are in a more mature state~\citep{Schuller14-CPE}.

A key challenge in the case of animal states and traits is that the ground truth remains elusive.
In the case of humans, it is possible to obtain the ground truth through self-evaluation~\citep{Schuller14-CPE}.
Moreover, it is also easier to rely on non-experts for \emph{annotation}, thus scaling up the amount of data that can be collected.
None of this is possible in bioacoustics, where only experts can properly interpret social cues and signals as well as connect vocalisations with states and traits.
A further challenge lies in the type of available data -- in the case of bioacoustics, this can be either too little or too much.
In the case of structured, well-defined experiments, it is hard to make `subjects' (i.e., animals) follow explicit instructions, resulting in little usable data, whereas with large-scale, passive monitoring, the data is often too much for humans to comprehend (e.\,g., the Australian acoustic observatory aims to collect 2 \emph{millenia} of sounds~\citep{Roe21-TAA}).

A middle-ground can be found by resorting to the monitoring of animals in captivity.
While captivity changes animal behaviour~\citep{McPhee04-GIC}, it results in a manageable quantity of data that can hopefully cover a suitable range of vocalisations and their underlying context.
Such datasets are now beginning to emerge in the public domain, raising the question: \emph{Can we predict animal states and traits in the same way that we can predict human ones?} 

We explore this question by utilising a recently open-sourced dataset of Egyptian fruit bats (\emph{Rousettus Aegyptiacus})~\citep{Prat17-AAD}.
This dataset contains a large number of \acp{USV} obtained in captivity, coupled with video information,
and has annotations for the context of each interaction -- for example, whether it takes place during \emph{feeding}, \emph{mating}, or \emph{fighting}.
This context implicitly encapsulates the social dynamics of the group and the current state of the addressee and emitter.
Correspondingly, it is expected to modulate bat \acp{USV} and be amenable to automatic prediction.
Follow-up work has shown that it is indeed possible to predict emitter, addressee, and context on a subset of the data~\citep{Prat16-EBV} using Mel-frequency cepstral coefficients and Gaussian mixture models -- but the entire range of available contexts has not been investigated yet.


While a large portion of previous work has focused on analysing the syntactic structure of \acp{USV}~\citep{Bohn08-SAT, Wang20-BIT, Amit23-BVS}, we approach our exploratory analysis from a ``paralinguistic'' perspective -- inspired by research on human vocalisations~\citep{Schuller14-CPE} -- and focus on pitch.
This follows a trend of recent work which investigates the roots of human vocalisations in mammalian ones~\citep{Portfors07-TAF, Brudzynski18-HOU}.
Specifically for bats, for instance, it has been shown that their high-intensity vocalisations exhibit similar patterns to humans, with an increase in rhythm and fundamental frequency~\citep{Bastian08-ACI}.
We thus begin with an exploratory analysis of the effect that context has on handcrafted pitch features derived from \acp{USV}.
We follow this up with classification experiments which investigate whether the automatic prediction of context is indeed possible, using both these pitch features and spectrogram-based \acp{DNN} -- this time using the full range of contexts on a larger dataset, compared to prior work~\citep{Prat16-EBV}.



\section{Dataset}
\label{sec:dataset}
We use a dataset of captive Egyptian fruit bats~\citep{Prat17-AAD}.
The dataset was collected by passive acoustic monitoring of multiple bats placed in two types of chambers: isolation chambers (mother and pup) and colony chambers (either adults before pregnancy or pups after weaning).
All chambers were continuously monitored with both infrared cameras and omnidirectional ultrasound microphones (Avisoft-Bioacoustics Knowles FG-O) recording at 250\,kHz.
The subjects were initially captured from a natural roost near Herzliya, Israel, and later placed in communal chambers.
Subsequently, a number of pups were born in captivity, with mothers placed in isolation before giving birth.
Pups were often isolated from mothers to test their feeding behaviour and considered mature once observed to feed by themselves.
Mother and pup were re-introduced to the colony immediately afterwards.

Most of the recordings took place between May 2012 and June 2013, with a follow-up recording period in February 2014.
The audio data was automatically segmented~\citep{Prat15-VLI} and subsequently manually annotated with the use of the infrared videos.
The annotations encompass the emitter and addressee, the pre- and post-vocalisation (re)actions of both, and the context in which the interaction took place.
Given space limitations, we focus exclusively on context, and thus ignore the other annotations.

Vocalisations were captured in all the setups described above.
The underlying context encompasses a total of 13 classes (with the total number of examples in parentheses): 
\emph{separation} (504) -- signifying the separation of an adult from the group;
\emph{biting} (1788) -- emitted by a bat after being bitten by another;
\emph{feeding} (6683) -- quarrelling over food;
\emph{fighting} (7963);
\emph{grooming} (383);
\emph{isolation} (5714) -- denoting the isolation of a pup from the group (immediately after birth and until adulthood);
\emph{kissing} (362);
\emph{landing} (16) -- emitted after one bat lands on top of another;
\emph{protesting} (2338) -- when a female protested a mating attempt;
\emph{threatening} (1065) -- aggressive displays without fighting;
\emph{general} (29627) -- all interactions of unspecified context, with bats generally keeping a longer distance than any other type of interaction;
\emph{sleeping} (33997) -- any aggressive interaction taking place in the sleep cluster.
We note that most annotated contexts pertain to acts of aggression where one bat violates the personal space of another, with the exception of grooming and kissing (and isolation/separation).
Additionally, the annotators specified the context of some interactions as \emph{unknown} (640) whenever they were unsure if it fits any of the above categories, which we excluded from further analysis.
Given its rare appearance, we also excluded \emph{landing}.
In some of the utterances it was impossible to identify who was the emitter and who was the addressee.
As this distinction was important for our analysis (especially the classification experiments; see below), we additionally excluded all utterances where the identity of the emitter was not clearly marked.
Finally, we removed utterances which lasted longer than 3 seconds as we consider them to be errors introduced in the annotation process (typical bat vocalisations last up to 1 second).
We ended with a total of 35\,074 vocalisations.

\section{Methodology}
In this section, we describe the methodology followed by our exploratory analysis and classification experiments.
We note that our code will be released upon acceptance to make our work reproducible.


\textbf{Fundamental frequency:}
We first extracted the \emph{fundamental frequency (F0)} for each vocalisation.
Here, we considered the entire utterance as our unit of analysis, and did not decompose it to syllables like prior work~\citep{Bastian08-ACI}.
This is more in line with the processing done for human speech, where F0 is estimated on longer chunks instead of individual phonemes or words~\citep{Schuller14-CPE}.
In detail, we computed the \ac{STFT} of the signal and then calculated the frequency which showed the maximum average energy over time;
all time-frequency bins with an amplitude lower than 20\,dB from this maximum energy were zeroed out to mitigate the influence of noise;
subsequently, we estimated the frequency with the maximum amplitude for each frame (manual inspection showed sufficient performance of this simplistic approach);
finally, we computed the following statistics of the resulting F0 contour: mean, standard deviation, maximum, minimum, and the linear slope fit on all F0 points;
this we did twice, once by taking into account silent frames, and once by excluding those from our statistics.
The \ac{STFT} was computed over 100\,ms windows with a hop size of 16\,ms.
We opted for longer windows that potentially encompass multiple syllables to obtain more robust estimates of F0, as we care about the general `prosody' of vocalisations.

\textbf{Classification experiments:}
We automatically classify the 11 classes of context that remain after excluding \emph{unknown} and \emph{landing}.
In general, we employ both a traditional, features-based classifier and a modern \ac{DNN}.

\textbf{Cross-validation:}
In all cases, we opted for an exhaustive 3-fold cross-validation, i.\,e., we made sure that each instance is used once (and only once) for testing.
We split the dataset into 3 folds, each featuring two subject-independent partitions, one for \emph{development} and one for \emph{testing}, taking care to keep the original balance of the labels in each one.
Ensuring that the development and testing splits do not feature the same individuals is essential to avoid information leakage, given that individual characteristics have been shown to influence mammalian vocalisations~\citep{Linhart22-TPF}.
Additionally, we further split each \emph{development} partition into \emph{training} and \emph{validation}; here, we opted for a simple 70-30\% stratified random split which did not take identity into account for simplicity.
The partitions will be released along with our code to foster reproducibility.

\textbf{Support vector machine (SVM):}
Our first experiment utilised a traditional, multiclass \textsc{SVM} using a one-vs-one setup.
As input, we used all 10 features described in the previous sub-section.
We explored values for the cost parameter of the \textsc{SVM} in $\{0.0001, 0.001, 0.005, 0.05, 0.1, 0.5, 1\}$.
The optimal parameter was chosen based on validation set performance separately for each fold (i.\,e., nested cross-validation); the model was then re-trained on the entire development set and evaluated on the testing partition.
The features were standardised based on the development set statistics of each fold.
To account for class imbalance, the cost parameter was further scaled with the inverse frequency of each class.
Our training framework was \textsc{scikit-learn-v1.1.3} in \textsc{python}.


\textbf{Deep learning:}
We further experimented with a \ac{DNN} architecture that operates on spectrograms -- a standard practice in computer audition research.
We opted for \textsc{ResNet50}~\citep{He16-DRL}, a standard architecture also shown to perform well for bioacoustics~\citep{Stowell22-CBW}, and \textsc{EfficientNet-B0}~\citep{Tan19-ERM}, the smallest member of the the \textsc{EfficientNet} family with approximately 4 Mio. parameters. 
In recent years, the \textsc{EfficientNet} family has been established as one of the most widely used \ac{CNN} architectures for the processing of images and audio spectrograms~\citep{Zeghidour21-LAL} in general, and for computational bioacoustic tasks in particular~\citep{Anderson22-LAF}. 
Its success can be particularly attributed its core \ac{CNN} components being scalable with a common factor.
This allows to easily explore different variants covering a large range of parameters.
Input spectrograms were computed with a window size of 4096 ($\approx$ 16\,ms, equal to the number of FFT bins, without padding) and a hop size of 10\,ms -- the shorter windows allow the spectrogram a higher temporal granularity than available for the pitch extractor.
To improve performance, we opted to initialise the model state with weights pre-trained on ImageNet -- this allows us to leverage the power of learnt representations and benefit from transfer learning, which has proven beneficial in prior work~\citep{Anderson22-LAF}.
For each fold, we trained a model for a maximum of 30 epochs using the \textsc{Adam} optimiser with a learning rate of 0.0001 and a batch size of 16, and finally picked the model state with the best performance on the validation set for evaluating on the test set.
To ensure a uniform data size, we zero-padded all sequences to the maximum of 3 seconds.
Models were trained using \textsc{torch-v1.13.0}.

\textbf{Performance:}
As seen in \cref{sec:dataset}, the dataset is heavily imbalanced.
We therefore opted for the use of \ac{UAR}\footnote{
UAR is also known as \emph{balanced accuracy}.
It is computed as the average recall per class.
We therefore opted for \ac{UAR} as term which best captures the way it is computed.
} 
as our evaluation metric, since it is designed to account for class imbalance.
To better gauge the robustness of our results, we collected the predictions for each fold and computed the \ac{UAR} on the entire dataset.
Additionally, we computed 95\% \acp{CI} using bootstrapping (where we randomly sample 1000 samples with replacement and compute the \ac{UAR} on those).

\section{Results \& Discussion}
\begin{table}[t]
    \centering
    \caption{
    F0 statistics for the different contexts, averaged over all vocalisations.
    Showing mean, standard deviation (std.), maximum, minimum, and linear slope.
    }
    \label{tab:pitch}
    \resizebox{\columnwidth}{!}{
    \begin{tabular}{lrrrrr}
\toprule
\textbf{Context} &  \textbf{Mean} [Hz] &  \textbf{Std.} [Hz] &  \textbf{Max} [Hz] &  \textbf{Min} [Hz] &  \textbf{Slope} [Hz]\\
\midrule
Biting (Bi)      &           11\,238 &           2198 &          15\,317 &           7761 &                9 \\
Feeding (Fe)     &           10\,584 &           2430 &          15\,264 &           6769 &              -29 \\
Fighting (Fi)    &           11\,680 &           2493 &          17\,563 &           7365 &                6 \\
General (Ge)     &           11\,098 &           2383 &          16\,220 &           7262 &                3 \\
Grooming (Gr)    &           11\,385 &           2212 &          15\,635 &           7919 &               35 \\
Isolation (Is)   &           12\,640 &           2150 &          17\,509 &           9684 &              -34 \\
Kissing (Ki)     &           11\,461 &           2128 &          15\,205 &           8162 &             -142 \\
Protesting (Pr)  &           11\,545 &           2702 &          17\,504 &           6877 &               15 \\
Separation (Se)  &           10\,196 &           2103 &          14\,658 &           6756 &               14 \\
Sleeping (Sl)    &           11\,500 &           2169 &          15\,715 &           7865 &              -15 \\
Threatening (Th) &           10\,838 &           2438 &          16\,581 &           6962 &               11 \\
\bottomrule
\end{tabular}}
\end{table}
\textbf{Exploratory data analysis} Corrected statistics of the fundamental frequency (i.\,e., excluding zeros) are shown in \cref{tab:pitch}.
The differences between contexts are overall low.
In our interpretation, we consider the \emph{general} context to be the equivalent of a \emph{neutral state} and all other states to deviate from that.
We note that the mean frequency across all contexts is approximately 11\,kHz and reaches a maximum of 17\,kHz and a minimum of 6\,kHz.
This is lower than results previously reported in the literature for bats~\citep{Gadziola12-SVO}, which might be caused by our simplistic computation of F0 (most other works compute pitch for individual syllables, whereas we compute it over the entire vocalisation).
Notably, the highest frequencies are observed during pup isolation and the lowest during adult separation -- this is expected as separation only appears for adult bats while isolation only for pups; with pups being smaller in size than adults, we expect them to reach higher pitches.
Moreover, adult separation shows a positive (rising) slope, whereas pup isolation a negative (falling) one -- an interesting finding suggesting that a more granular analysis of the pitch contours may yield better predictive performance.

\begin{table}[t]
    \centering
    \caption{UAR and 95\% CIs on the entire dataset.}
    \label{tab:results}
    \begin{tabular}{c|c}
        \toprule
        \textbf{Model} & \textbf{UAR} \\
        \midrule
        \textsc{SVM} & .224 [.213 - .234] \\
        \textsc{ResNet50} & .277 [.270 - .284]\\
        \textsc{EfficientNet} & \textbf{.333 [.325 - .340]}\\
        \bottomrule
    \end{tabular}
\end{table}

\begin{figure}[t]
    \centering
    \includegraphics[width=\columnwidth]{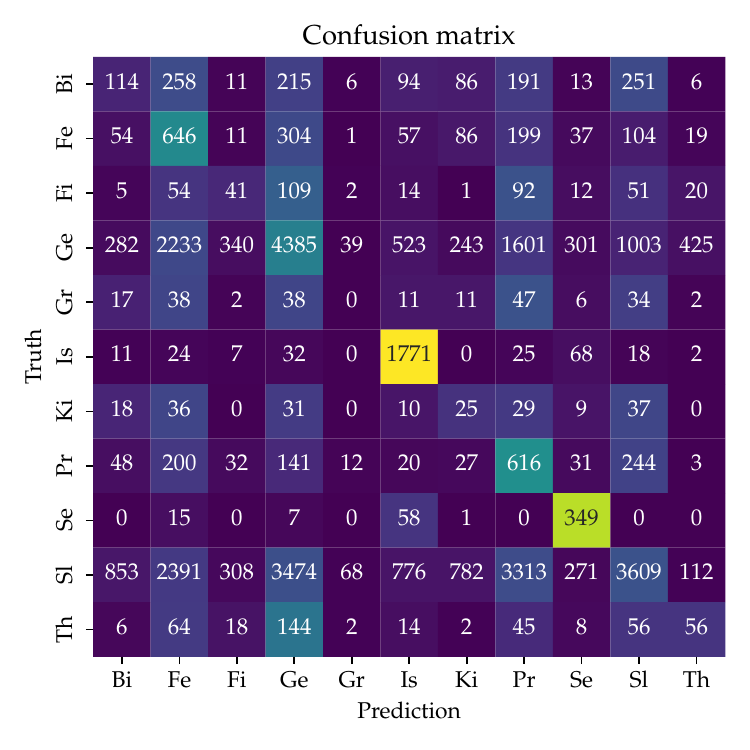}
    \caption{
    Confusion matrix for the best-performing model.
    Classes:
    Bi $\rightarrow$ biting;
    Fe $\rightarrow$ feeding;
    Fi $\rightarrow$ fighting;
    Ge $\rightarrow$ general;
    Gr $\rightarrow$ grooming;
    Is $\rightarrow$ pup isolation;
    Ki $\rightarrow$ kissing;
    Pr $\rightarrow$ protesting mating;
    Se $\rightarrow$ adult separation;
    Sl $\rightarrow$ sleeping;
    Th $\rightarrow$ threatening.
    Results colour-coded per row.
    }
    \label{fig:confusion}
\end{figure} 

\textbf{Classification experiments} Our results are shown in \cref{tab:results}.
We obtain $22.4\%$ when using our simple features and an \textsc{SVM} and $33.3\%$ for \textsc{EfficientNet}.
Results illustrate that the classification of context from vocalisations is indeed possible as chance-level is $\approx 9\%$ for 11 classes.
They further show how \acp{DNN} can leverage pre-trained representations to obtain better performance compared to handcrafted features.

\textbf{Discussion} Additionally, \cref{fig:confusion} shows the confusion matrix for the best-performing model (\textsc{EfficientNet}).
This gives additional insight into the types of confusions made by the model.
We observe that several samples are mislabelled as \emph{sleeping}.
This is expected given that this class is overrepresented in our data.
Additionally, \emph{sleeping} is also often mistaken as other classes, despite being over-represented, an indication that the boundaries between classes might not be entirely distinct.
On the other hand, \emph{grooming} shows a recall of 0, which is also expected as it is one of the two least represented classes.
\emph{Protesting} and \emph{feeding} are also drawing a lot of misclassifications from general and sleeping; this might indicate annotation ambiguity in the case where interactions took place in the sleeping context but were related to the first two activities, or where the interaction was not strong enough for the annotators to observe.

\emph{General} and \emph{sleeping} are also often confused as pup isolation.
This could be a side-effect of the overrepresentation of the first two classes, but may also be indicative of pups continuing to emanate isolation vocalisations even after being integrated with the main colony -- i.\,e., a side-effect of delayed social adaptation which is known to take place for this species~\citep{Prat15-VLI}.
Interestingly, \emph{pup isolation} and \emph{adult separation} are most confused with one another; on the one hand, this might indicate that these two types of vocalisations are most similar; however, an alternative interpretation is that as pups progressed to the adult stage, their vocalisations gradually transitioned, and that the hard cutoff the original authors have used might have left some edge cases which lead to confusion; finally, as both classes were collected in the isolation chambers, while all others in the colony chambers, this similarity might simply be a side-effect of different background acoustics.
We further note that this discrepancy is not reflected in the F0 statistics shown in \cref{tab:pitch}; there, isolation showed the highest F0 and separation the lowest.
This illustrates how feature-based analyses are perhaps ill-suited for the intricacies of \acp{USV}, which can be uncovered through the use of \acp{DNN}.

Generally, as the dataset has been annotated in great detail regarding the times of birth for each new pup and its subsequent introduction to a group chamber, we expect further insights to come from a follow-up, post-hoc analysis of errors.
These could clarify whether the misclassifications we describe above are simply a side-effect of underperformance, or whether the underlying cause is the presence of subclasses hidden within the annotations (e.\,g., the gradual transition of pups into adulthood).

\section{Conclusion}
We conducted an analysis of the interaction context in bat \acp{USV} for a cohort of Egyptian fruit bat subjects exhaustively monitored over a multi-month period in captivity.
Statistical analysis of the fundamental frequency showed small differences across the different interaction contexts.
Nevertheless, spectrogram-based \acp{DNN} could predict context with a \ac{UAR} of $33.3\%$ for 11 classes -- more than three times chance-level ($9\%$).
Our analysis further indicates that the confusions between different classes may map to underlying hidden factors (such as social adaptation after introducing a pup to a colony).
Future work may further investigate these nuances by analysing the interplay of the different factors over time -- a context in which handcrafted features may indeed prove more useful.
Broadly, our work showcases the feasibility and promise of analysing animal speaker states and potentially also traits from their vocalisation.
Most excitingly, this line of research may pave the way for the automatic analysis of animal health status -- a topic of fundamental importance for epidemiology~\citep{Johnson23-AAA}.

\section{\refname}
\printbibliography[heading=none]
\end{document}